\documentclass[10pt,aps,prd,twocolumn,notitlepage,nofootinbib,floatfix,unsortedaddress,preprintnumbers]{revtex4-1}

\usepackage{amsmath, amsthm, amssymb, amsfonts, amsbsy, mathrsfs}

\usepackage{multirow}

\begin{document}
\preprint{YITP-25-65, IPMU25-0019}

\title{
    Emergent Lorentzian dispersion relations from a Euclidean
    scalar-tensor theory
    }

\author{Justin C. Feng}
\email{feng@fzu.cz}
\affiliation{
    Central European Institute for Cosmology and Fundamental Physics
    (CEICO), Institute of Physics of the Czech Academy of Sciences, Na
    Slovance 1999/2, 182 21 Prague 8, Czech Republic
    }

\author{Shinji Mukohyama}
\affiliation{
    Center for Gravitational Physics and Quantum Information (CGPQI),
    Yukawa Institute for Theoretical Physics (YITP),\\ Kyoto University,
    Kyoto 606-8502, Japan} \affiliation{Kavli Institute for the Physics
    and Mathematics of the Universe (WPI), \\The University of Tokyo
    Institutes for Advanced Study, \\The University of Tokyo, Kashiwa,
    Chiba 277-8583, Japan
    }

\author{Sante Carloni}
\affiliation{
    Institute of Theoretical Physics, Faculty of Mathematics and
    Physics, Charles University, Prague, V Hole\v sovi\v ck\'ach 2, 180
    00 Prague 8, Czech Republic
    }
\affiliation{
    DIME, Università di Genova, Via all'Opera Pia 15, 16145 Genova,
    Italy
    }
\affiliation{
    INFN Sezione di Genova, Via Dodecaneso 33, 16146 Genova, Italy
    }



%
%

%
%
\begin{abstract}
Can one be fooled into thinking that space and time are fundamentally described by a Lorentzian manifold? In this article, we describe a scenario in which a theory constructed on a (Euclidean signature) Riemannian manifold can lead to degrees of freedom with Lorentzian dispersion relations, due to a non-trivial configuration of a scalar field. In particular, we perform a perturbative analysis of a renormalizable shift-symmetric scalar-tensor theory and find that it can, in principle, admit a massless tensor degree of freedom with a Lorentzian dispersion relation. While the remaining degrees of freedom in the gravity sector will, in general, satisfy Euclidean dispersion relations, we argue that they can be brought under control by elliptic equations with an appropriate choice of boundary conditions.
\end{abstract}


%
%

\maketitle

%
%

%
%

\section{Introduction}

In physics, the dynamics of a given system is nothing but a sequence of configurations parameterized by time. For example, in general relativity, the dynamics of the universe is represented by the spacetime itself, which can be considered as a sequence of spatial configurations parameterized by time, up to general coordinate transformations. However, as we go back towards the initial singularity, the geometrical description of the universe should break down, and thus, space may be emergent. Then a natural question is whether time can also be emergent. 

The idea that time may be emergent is in line with the fact that in any diffeomorphism invariant theory of gravity, the Hamiltonian is a linear combination of constraints and thus vanishes up to boundary terms, meaning that there is no evolution of quantum states in the bulk; this issue is sometimes referred to in the literature as the problem of time (see the reviews \cite{Anderson:2012vk,*Isham:1992ms}). It is usually thought that the dynamics should then be encoded as correlations among various fields. In other words, one of the fields should play the role of time, one such example being an inflaton field during inflation. In this sense, the concepts of time and dynamics may be emergent. 

Motivated by the above set of thoughts, ref.~\cite{Mukohyama:2013ew} proposed a scenario in which the concept of time in the sense of the Lorentz signature of an effective metric emerges from a locally Euclidean theory without time. The basic idea is to introduce a clock field, i.e., a field playing the role of time. Since it must carry at least one number corresponding to the time, the simplest possible candidate for the clock field is a scalar field, which was adopted in the proposal. Since we would not like to induce violation of the time translation and/or reflection symmetries in the matter sector from the gravity sector after the emergence of time, we demand that the gravity action including the clock field $\varphi$ is invariant under the shift symmetry, $\varphi\to \varphi+\mbox{const}$, and the $Z_2$ symmetry, $\varphi\to -\varphi$. When the clock field $\varphi$ develops a sufficiently large gradient in a region of the Riemannian, i.e., locally Euclidean, manifold, the concept of time emerges along the direction of the gradient. 

Mathematically, this phenomenon can be described as the signature change of an effective metric, which is a disformal transformation \cite{Ishak:2018his,Bekenstein:1992pj} of the originally positive-definite metric by the clock field $\varphi$. In regions with a small or vanishing gradient of the clock field, the effective metric remains positive definite, and the perturbations are described by elliptic equations. On the other hand, in regions with a large gradient of the clock field, the effective metric becomes Lorentzian, and the perturbations with low momenta are well described by hyperbolic equations. 

Hence, the signature change takes place through a hypersurface on which the effective metric becomes degenerate. Near the boundary, higher derivative terms should become important and the description of the system at the leading order in the derivative expansion completely breaks down.\footnote{Signature change has been the subject of heated debate in the past; see \cite{White:2008xr} and the papers and comments listed in their refs. [14-15] for a survey of the debate regarding signature change in classical general relativity.} Even away from the boundary, the description of short distance behavior of the system requires inclusion of higher derivative terms. For these reasons, ref.~\cite{Mukohyama:2013gra} proposed a renormalizable higher-derivative theory of Riemannian gravity with a clock field as a possible UV completion. 

The purpose of the present paper is to study perturbations around a simple background solution with a non-vanishing and constant derivative of the clock field in the renormalizable Riemannian theory proposed in ref.~\cite{Mukohyama:2013gra}.  We show that for a range of parameters, gravitational waves at long distances indeed have a positive time kinetic term and follow a hyperbolic dispersion relation. 

The rest of the paper is organized as follows. In section \ref{sec:Theory}, we review the theory in \cite{Mukohyama:2013gra} and present a reformulation suitable for our analysis. In section \ref{sec:Perturbation}, we construct perturbations about a flat background and present expressions for the resulting Lagrangian (keeping terms quadratic in the perturbations). In section \ref{sec:ScalarSector}, we analyze in detail the sector containing scalar degrees of freedom. Finally, we summarize and discuss our results in section \ref{sec:Disc}.

%
%
\section{Theory} \label{sec:Theory}
\subsection{A shift-symmetric emergent Lorentz signature theory}
A general shift symmetric Emergent Lorentz Signature Theory (ELST) with terms containing no more than four
derivatives of the fields is given by the action \cite{Mukohyama:2013gra}:
\begin{equation} \label{Eq:ELSTAction}
    \begin{aligned}
    S = \int_M d^4x \sqrt{|g|} L, \qquad L:=L_0+L_2+L_4,
    \end{aligned}
\end{equation}
where:
\begin{equation} \label{Eq:ELSTLagrangian}
    \begin{aligned}
    L_0 &:= c_{11}, \qquad\qquad\qquad\qquad\quad\>\>
    L_2 := c_{9} R + c_{10} X , \\
    L_4 &:= 
        c_1 R^2 + c_2 R_{ab} R^{ab} + c_3 R_{abcd} R^{abcd} + c_4 X R\\
        & \quad \>
        + c_5 R^{ab}\varphi_a \varphi_b + c_6 {\rm X}^2
        + c_7 (\Box \varphi)^2+ c_8 \varphi_{ab} \varphi^{ab} .
    \end{aligned}
\end{equation}
with the following definitions for the derivatives of the clock field $\varphi$:
\begin{equation} \label{Eq:Defs}
  \varphi_a:=\nabla_a\varphi,
  \qquad
  \varphi_{ab}:=\nabla_a\nabla_b\varphi,
  \qquad
  X:= \varphi^a \varphi_a.
\end{equation}
This theory has been shown to be renormalizable \cite{Muneyuki:2013aba} (which is expected, as an action containing quadratic curvature terms is renormalizable \cite{Stelle:1976gc}), and since the theory has Euclidean signature, one can choose the couplings so that the action is bounded below \cite{Mukohyama:2013gra}. Moreover, it was recently shown \cite{Feng:2023klt} that this ELST can in principle provide a regular description for the quasiregular singularities that are necessarily present at the termination of the horizon for evaporating black holes in the ``baby universe'' scenario \cite{Hossenfelder:2009xq,*Hawking:1990jb,*Hawking:1988wm,*Hawking:1988ae}.

In \cite{Mukohyama:2013gra,Mukohyama:2013ew}, the emergence of Lorentz signature arises from matter couplings; in
particular, one presumes that (at long-distance scales\footnote{We will make this notion more precise later, but here we mean regimes in which higher derivative contributions to the action can be neglected.}) matter is
coupled to an effective metric:
\begin{equation} \label{Eq:EffectiveMetric}
    {\bf{g}}_{ab} = g_{ab} - 
                        \frac{\varphi_a  \varphi_b}
                        {X_C},
                        \qquad
    \bar{\bf{g}}^{ab} = g^{ab} - 
                        \frac{\varphi^a \varphi^b}
                        {X-X_C},
\end{equation}
where $X_C$ is a positive constant that appears in the matter couplings
and $X:= \varphi^a \varphi_a$, and we employ the convention that indices
are raised and lowered with the Euclidean metric $g_{ab}$ and its
inverse $g^{ab}$. Whether this can be realized will require a careful
study of the renormalization group flow in the matter sector (with the 
idea of the emergent Lorentz invariance as in \cite{Chadha:1982qq}); this is an
important question which is left for future work. The main purpose of this article is to consider the
theory in the absence of matter, and to ask whether 
a gravitational theory formulated on a Euclidean signature manifold can, in the weak field long distance limit, mimic the behavior of
gravitational degrees of freedom propagating on a Lorentzian manifold.

\subsection{Reformulation of action}

It is perhaps convenient to reformulate the action. With the parameter
relations:
\begin{equation} \label{Eq:Parrel}
    \begin{aligned}
    c_1 &= -\frac{\eta}{4}+\frac{1}{6\lambda}+\sigma, 
    & 
    c_2  = -\frac{1}{\lambda}-4\sigma, 
        \quad\>\>\, &
    c_3  = \frac{1}{2\lambda}+\sigma, \\
    c_4 &= -\gamma_0/2, 
        \qquad & 
    c_5  = \gamma_0, 
        \qquad\qquad\>\> &
    c_6  = 1, \\
    c_7 &= \alpha_0/2, 
        \qquad & 
    c_8  = \beta_0, 
        \qquad\qquad\>\> &
    c_9  = -Z, \\
    c_{10} &= -2 X_0, 
    \quad & 
    c_{11}  = P_0+X_0^2,\quad\>\,
    \end{aligned}
\end{equation}
and integration by parts, one may write the dynamically equivalent action $S':=\int_M
d^4x\sqrt{|g|}\tilde{L}$, where $\tilde{L}$ is given by:\footnote{The notation and conventions employed here differ slightly from that of Eq. (2.5) of \cite{Mukohyama:2013gra}. Here, we employ abstract index notation \cite{Wald,*HawkingEllis}, with indices raised and lowered using the Euclidean signature metric $g_{ab}$. For later convenience, we also: 1) rewrite the $R^2$ term in (2.5) in terms of a scalar quantity $\chi$, 2) modify the cosmological term so that terms involving only $X_0$ and $X$ form a complete square, and 3) we subtract boundary terms so that the $R^{ab} \varphi_a \varphi_b$ term may be combined with $\varphi_a \varphi^a$ to form the $G^{ab} \varphi_a \varphi_b$ term.}
\begin{equation} \label{Eq:ELSTLagrangianRF}
    \begin{aligned}
    \tilde{L} &:= P_0 + X_0^2 - Z R + \eta (\chi^2- \chi R) + \frac{C_{abcd}C^{abcd}}{2\lambda} - 2X_0 X \\
    &\quad + X^2 + \gamma_0 G^{ab} \varphi_a \varphi_b + \alpha_0 \varphi{_a}{^a} \varphi{_b}{^b} + \beta_0 \varphi^{ab} \varphi_{ab}+\sigma E,
    \end{aligned}
\end{equation}
where $C_{abcd}$ is the Weyl tensor, $\chi$ is an auxiliary field
(replacing a term quadratic in the Ricci scalar), and $E$ is the
Gauss-Bonnet invariant satisfying:
\begin{equation} \label{Eq:GaussBonnet}
    \begin{aligned}
    E 
    =&~C_{abcd}C^{abcd}-2R_{ab}R^{ab}+\tfrac{2}{3}R^2.
    \end{aligned}
\end{equation}

To recover the appropriate flat space solution with $X=X_0$ (corresponding to a flat spacetime background), we require that $P_0=0$. The constants $Z$ and $X_0$ are dimensionful; we assume that they are on the order of the Planck mass squared $M_\mathrm{Pl}^2$. 

\section{Perturbation about a flat background} \label{sec:Perturbation}

\subsection{Background and perturbation}

We then consider the perturbation of $\tilde{L}$ about
the flat background satisfying (setting $P_0=0$):
\begin{equation} \label{Eq:BackgroundMetricGradient}
    \bar{g}_{ab} = \text{diag}(1,1,1,1), \quad \bar\varphi=\sqrt{X_0}t, \quad \bar\chi=0.
\end{equation}
with coordinates $(t,x,y,z)$. The perturbations are written in terms of spatial Fourier modes with wavenumber $\vec{k}$ in the direction of the coordinate $x$. With this in mind, we consider the following form for the metric perturbation (where $s$ is the
expansion parameter):
\begin{equation} \label{Eq:PerturbedMetric}
    \delta{g}_{ab} = 
        s
        \left[
        \begin{tabular}{cccc}
            $U$     &   0   &    $B_y$   & $B_z$      \\
            0     &   $\psi$   &    0       & 0          \\
            $B_y$ &   0   & $\psi+h_+$ & $h_\times$ \\
            $B_y$ &   0   & $h_\times$ & $\psi-h_+$   
        \end{tabular}
        \right],
\end{equation}
in which we decompose the metric perturbations into tensor ($h_+$, $h_\times$), vector ($B_y$, $B_z$), and scalar ($U$, $\psi$) components, classified according to their behavior under spatial rotations about the $x$-axis. We also consider perturbations of the scalar fields:
\begin{equation} \label{Eq:BackgroundScalar}
    \delta{\varphi} = s \phi, \qquad \delta \chi = s \xi.
\end{equation}
For each quantity $U$, $\psi$, $\phi$, $\xi$, $B_{I}$, $h_{K}$ (with
$I\in\{y,z\}$, $K\in\{+,\times\}$) in the perturbation, we pick out a
spatial mode of the form:
\begin{equation} \label{Eq:ModeDecomp}
    \begin{aligned}
    F &= F_1(t) \sin(k x) + F_2(t) \cos(k x).
    \end{aligned}
\end{equation}
We expand $S'$ to second order in the expansion parameter $s$ to obtain
an action for small amplitude perturbations, which may be cleanly
separated into a tensor, vector, and scalar part.

\subsection{Tensor sector}
After performing the mode decomposition and integrating over the spatial coordinates, the action for the tensor sector
(containing $h_\times$ and $h_+$) consists of terms of the form (terms of order $O(s)$ cancel out when setting $P_0=0$):
\begin{equation} \label{Eq:PerturbedActionTensorSector}
    \begin{aligned}
    \underline{S}_h &= \frac{s^2 V_0}{8} \int dt \biggl[\{X_0 (\beta_0+\gamma_0)+Z\}\dot{h}^2-(\gamma_0 X_0-Z)k^2 h^2\\
    &\qquad\qquad\qquad+\frac{1}{\lambda}\left(4 k^2 \dot{h}^2+\mathcal{T}^2\right)\biggr],
    \end{aligned}
\end{equation}
where $V_0$ is a volume factor, and $\mathcal{T}$ is defined as:
\begin{equation} \label{Eq:HigherDerivTensor}
    \begin{aligned}
    \mathcal{T} &:= \ddot{h}+k^2 h.
    \end{aligned}
\end{equation}
Now $X_0$ and $Z$ both have mass dimension two, so that we can characterize the 
long-distance limit as $k \ll M_{\rm Pl}$, $\dot h \ll M_{\rm Pl}$, the
terms proportional to $X_0$ and $Z$ dominate. We require that
\begin{equation} \label{Eq:ConditionGhostFree0}
    \begin{aligned}
    (X_0 (\beta_0+\gamma_0)+Z)>0,
    \end{aligned}
\end{equation}
so that the tensor modes have a positive Kinetic term and also that
\begin{equation} \label{Eq:ConditionLorentzianTensor}
    \begin{aligned}
    (X_0 (\beta_0+\gamma_0)+Z)(\gamma_0 X_0-Z)>0,
    \end{aligned}
\end{equation}
so the perturbations $h$ will satisfy a massless Lorentzian dispersion
relation in the long-distance limit; the tensor degrees of freedom will
therefore appear as massless fields propagating on a Lorentzian background. 

\subsection{Vector sector}
On the other hand, the sector containing the vector degrees of freedom
$B_x$ and $B_y$ consist of terms of the form:
\begin{equation} \label{Eq:PerturbedActionVectorSector}
    \begin{aligned}
        \underline{S}_B=
        \frac{s^2 V_0 k^2}{4 \lambda} \int dt 
        \biggl[
        \dot{B}^2+ k^2 B^2 + \lambda  (X_0 (\beta_0+\gamma_0)+Z) B^2
        \biggr].
\end{aligned}
\end{equation}
The resulting dispersion relation is Euclidean rather than Lorentzian
and it contains a large mass term that is tachyonic relative to $\dot{B}^2$ if all the constants are assumed to be positive. However, a Euclidean dispersion relation, which one might expect to arise for modes governed by elliptic equations, indicates that the modes are determined by a suitable boundary condition. If one chooses, e.g., $B=0$ at the boundary, a large tachyonic mass will suppress these modes in the bulk.

\subsection{Scalar sector}
The scalar sector containing the fields $(U,\psi,\phi,\xi)$ is
more complicated. It is convenient to adopt an approach outlined in \cite{Mukohyama:2014rca} and \cite{Aoki:2019snr} (which resembles the spin projection operator formalism \cite{Barker:2024juc,Buoninfante:2016iuf,Lin:2018awc,*Lin:2020,Aurilia:1969bg}), in which the quadratic action is represented in terms of vectors (containing the degrees of freedom) and matrices containing information about the Kinetic terms, the potential, and friction terms on the second derivatives $\ddot\phi$, so one cannot immediately express the quadratic action in first order form. Following \cite{Aoki:2019snr}, one may add to the scalar sector Lagrangian $L_\Phi$ a Lagrange multiplier term of the form $\Theta (\sqrt{2}\Omega-\Box \phi)$, where $\Theta$ is a Lagrange multiplier and $\Omega$ is a new variable, and redefine the Lagrange multiplier $\Theta$ to replace $\Box\phi$ in the quadratic action with $\Omega/2\sqrt{2}$. One may then integrate by parts to convert terms of the form $\Theta \ddot\phi$ to the first order form $\dot \Theta \dot \phi$. After performing this procedure and dropping surface terms, the resulting scalar action may be written as:
\begin{equation} \label{Eq:PerturbedActionScalarSector}
    \begin{aligned}
        \underline{S}_\Phi=
        s^2 V_0 \int dt 
        \underline{L}_\Phi.
\end{aligned}
\end{equation}
where the Lagrangian $\underline{L}_\Phi$ has the quadratic form (following the general decomposition in \cite{Mukohyama:2018obj}):
\begin{equation} \label{Eq:PerturbedLagrangianScalarSectorForm}
    \begin{aligned}
        \underline{L}_\Phi=&
            \dot{Y}^T \cdot \underline{\mathbb{K}} \cdot \dot{Y}
            +
            {Y}^T \cdot \underline{\mathbb{V}} \cdot {Y}
            +
            \dot{Y}^T \cdot \underline{\mathbb{M}}^T \cdot {Y}\\
            &
            +
            {Y}^T \cdot \underline{\mathbb{M}} \cdot \dot{Y}
            +
            {Y}^T \cdot \mathbb{A}^T \cdot {Z}
            +
            {Z}^T \cdot \mathbb{A} \cdot {Y}\\
            &
            +
            \dot{Y}^T \cdot \mathbb{B}^T \cdot {Z}
            +
            {Z}^T \cdot \mathbb{B} \cdot \dot{Y}
            +
            {Z}^T \cdot \mathbb{C} \cdot {Z}.
    \end{aligned}
\end{equation}
The variables are encoded in the vectors:
\begin{equation} \label{Eq:PerturbedLagrangianScalarSector}
    \begin{aligned}
        {Y}=& (\psi,\xi,\phi,\Theta), \qquad {Z}=(\Omega,U),
\end{aligned}
\end{equation}
where $Y$ contains the dynamical variables and $Z$ are auxiliary fields.

\begin{table*}[]
    \center
    \caption{Constants and $k$-dependent coefficients}
    \begin{tabular*}{0.92\textwidth}{|l|l|l|l|l}
        \cline{1-3} \cline{5-5}
        \textbf{Dimensionless} & \multicolumn{2}{l|}{\textbf{Dimensionful}} & \multirow{5}{*}{$\quad$} & \multicolumn{1}{l|}{\textbf{k-dependent}} \\[2pt]
        \cline{1-3} \cline{5-5}
        \cline{1-3} \cline{5-5}
        $\bar\nu_{\rm (a,b,c)}={\rm a}\alpha_0+{\rm b}\beta_0+{\rm c}\gamma_0$ & $\bar\mu_{{\rm (a,b,c,d)}}=X_0 \bar\nu_{\rm (a,b,c)}+Z {\rm d}$ & $\mathcal{K}=-3\mu_4/(2\nu_1)$ & \multicolumn{1}{l|}{} & \multicolumn{1}{l|}{$\kappa_1=3\mu_{1}\lambda +k^2$} \\[2pt] \cline{1-3} \cline{5-5}
        $\nu_1=\alpha_0+\beta_0$ & $\mu_1=X_0 \gamma_0 + Z$ & $\mu_4=3 X_0 (\gamma_0-2\beta_0)+Z$ & \multicolumn{1}{l|}{} & \multicolumn{1}{l|}{$\kappa_2=3\mu_{2}\lambda +k^2$} \\[2pt] \cline{1-3} \cline{5-5}
        $\nu_2=2\alpha_0+\beta_0$ & $\mu_2=X_0 \gamma_0 - Z$ & $\mu_5=3 X_0 \beta_0^2 + 2 \nu_1 \mu_4$ & \multicolumn{1}{l|}{} & \multicolumn{1}{l|}{$\kappa_3=2X_0 +k^2\beta_0$} \\[2pt]\cline{1-3} \cline{5-5}
        $\nu_3=3\alpha_0+\beta_0$ & $\mu_3=X_0 (\beta_0+\gamma_0)+Z$ & $\mu_6=3X_0(\beta_0^2+4\eta)+2\mu_4\nu_1$ & \multicolumn{1}{l|}{} & \multicolumn{1}{l|}{$\kappa_4=3\kappa_{3}X_0\lambda+k^4$} \\[2pt]
        \cline{1-3} \cline{5-5}
        &  \multicolumn{2}{l|}{$\mu_7=\bar\mu_{(3,1,-2,-2)}=X_0(3\alpha_0+\beta_0-2\gamma_0)-2 Z$}  & \multicolumn{1}{l}{} & \\[2pt]
        \cline{1-5}
        & \multicolumn{4}{l|}{$\mu_6=18 X_0 (4\eta-\alpha_0\beta_0)+\nu_1[12\mu_1-2\bar\mu_{(0,4,-1,-3)}+3\eta \lambda(\mu_2+10\mu_3)]$} \\[2pt]
        \cline{1-5}
        & \multicolumn{4}{l|}{$\mu_s^2=4 X_0 (3\eta \lambda \bar\mu_{(0,4,5,3)} -2 \bar\mu_{(3,1,-1,-3)})+3\lambda \left[4\mu_1^2\nu_1-X_0 \beta_0(4 X_0\alpha_0\gamma_0+\beta_0 \bar\mu_{(6,2,-1,-3)})\right]\>\>\>\,$} \\[2pt]
        \cline{1-5}
    \end{tabular*}
    \label{Tab:constants_and_variables}

    {The dimensionful constants $(\mu_i,\mathcal{K})$ have mass dimension two and are expected to be of $\mathcal{O}(M_{\rm Pl}^2)$.}
\end{table*}

\begin{table*}[htp]
    \centering
    \caption{Matrix coefficients for ${L}_\Phi$}
    \begin{tabular*}{0.92\textwidth}{@{\extracolsep{\fill}}|l|l|l|}
        \hline
        \textbf{Kinetic matrix $\mathbb{K}$ coeffs.}  & \textbf{Mass matrix $\mathbb{V}$ coefficients} & \textbf{Friction matrix $\mathbb{M}$ coefficients}  \\
        \hline
        \hline
        ${K}_{11} = \mathcal{K} \qquad\quad\> {K}_{12} = -\frac{3\eta}{4\sqrt{2}} \quad $ 
        & ${V}_{11} = \frac{k^2(\kappa_2\kappa_4-k^2\kappa_1^2)}{3\kappa_4 \lambda} \quad\>\> {V}_{12} = -\frac{k^2\eta(2\kappa_4+k^2\kappa_1)}{4\sqrt{2}\kappa_4}\>\>\>$ 
        & ${M}_{13} = \frac{\sqrt{X_0} k^2 \left[\beta_0 \kappa_4(4\nu_1-3\beta_0)-2\kappa_1\kappa_3\nu_1\right]}{2 \kappa_4\nu_1}$ 
        \\
        ${K}_{33} = \frac{k^4 \kappa_3}{\kappa_4} \qquad {K}_{34} = \frac{k^4 }{2 \sqrt{2} \kappa_4}$ 
        & ${V}_{22} = \frac{\eta(2\kappa_4-3k^4\eta\lambda)}{32 \kappa_4}$ 
        & ${M}_{14} = -\frac{\sqrt{X_0} \left(3 \beta_0\kappa_4+(k^2\kappa_1-3\kappa_4)\nu_1\right)}{2 \sqrt{2} \kappa_4\kappa_1}$ 
        \\
        ${K}_{44} = - \frac{3 X_0 \lambda}{8 \kappa_4}$ 
        & ${V}_{33} = -\frac{k^4 \beta_0\nu_2}{2 \nu_1}
        \qquad\quad\>\>\> {V}_{34} = \frac{k^2\alpha_0}{2\sqrt{2} \nu_1}$ 
        & ${M}_{23} = -\frac{3 \eta \lambda \sqrt{X_0} k^2 \kappa_3}{4 \sqrt{2} \kappa_4}$ 
        \\
        $\,$ 
        & ${V}_{44} = -\frac{1}{4 \nu_1}$ 
        & ${M}_{24} = -\frac{3 \eta \lambda \sqrt{X_0} k^2 }{16 \kappa_4}$ 
        \\
        \hline
    \end{tabular*}
    \label{Tab:MatrixCoefficients}
\end{table*}

Before listing the components of the matrices $\mathbb{K}$,
$\mathbb{V}$, $\mathbb{M}$, $\mathbb{A}$, $\mathbb{B}$, $\mathbb{C}$, it
is perhaps convenient to define a set of constants and $k$-dependent
coefficients. These definitions are listed in Table
\ref{Tab:constants_and_variables}; we will refer to these definitions
throughout. In terms of the definitions given in Table
\ref{Tab:constants_and_variables}, the components of the
$4\times4$ Kinetic matrix $\underline{\mathbb{K}}$ are:
\begin{equation} \label{LagMatK}
    \begin{aligned}
    \underline{\mathbb{K}}&=
    \left[
    \begin{tabular}{cccc}
        $-3 \mu_4$ & $-\frac{3 \eta }{4 \sqrt{2}}$ & 0 & 0 \\
        $-\frac{3 \eta }{4 \sqrt{2}}$ & 0 & 0 & 0 \\
        0 & 0 & $\kappa_3$ & $\frac{1}{2 \sqrt{2}}$ \\
        0 & 0 & $\frac{1}{2 \sqrt{2}}$ & 0  
    \end{tabular}
    \right],
\end{aligned}
\end{equation}
the $4\times4$ matrix $\underline{\mathbb{V}}$ has the components:
\begin{equation} \label{LagMatV}
    \begin{aligned}
    \underline{\mathbb{V}}=
    \left[
    \begin{tabular}{cccc}
        $\frac{k^2 \kappa_2}{3 \lambda }$ & $-\frac{\eta  k^2}{2 \sqrt{2}}$ & 0 & 0 \\
        $-\frac{\eta  k^2}{2 \sqrt{2}}$ & $ \frac{\eta}{16}$ & 0 & 0 \\
        0 & 0 & $k^4 \beta_0$ & $\frac{k^2}{2 \sqrt{2}}$ \\
        0 & 0 & $\frac{k^2}{2 \sqrt{2}}$ & 0 
    \end{tabular}
    \right],
\end{aligned}
\end{equation}
and the $4\times4$ friction matrix $\mathbb{M}$ has components:
\begin{equation} \label{LagMatM}
    \begin{aligned}
    \underline{\mathbb{M}}=
    \sqrt{X_0}\left[
    \begin{tabular}{cccc}
        0 & 0 & $2 k^2  \beta_0$ & $\frac{3}{2 \sqrt{2}}$ \\
        0 & 0 & 0 & 0 \\
        0 & 0 & 0 & 0 \\
        0 & 0 & 0 & 0 
    \end{tabular}
    \right],
\end{aligned}
\end{equation}
Note that $\mathbb{K}$ and $\mathbb{V}$ are symmetric, but $\mathbb{M}$
is not symmetric. The matrices $\mathbb{A}$, $\mathbb{B}$, and
$\mathbb{C}$ have the form:
\begin{equation} \label{LagToyModelMatxs}
    \begin{aligned}
    &\mathbb{A}=
    \left[
    \begin{tabular}{cccc}
        0  & 0   & $\frac{\beta_0 k^2}{\sqrt{2}}$ & $\frac{1}{2}$   \\
        $-\frac{k^2 \kappa_1}{6 \lambda }$ & $-\frac{\eta  k^2}{8 \sqrt{2}}$ & 0 & 0 
    \end{tabular}
    \right]
    ,\qquad
    \mathbb{C}=
    \left[
    \begin{tabular}{cc}
        $\nu_1$  & 0     \\
        0       & $\frac{\kappa_4}{12 \lambda }$
    \end{tabular}
    \right]
    ,\\
    &\qquad\>
    \mathbb{B}=
    \sqrt{X_0}
    \left[
    \begin{tabular}{cccc}
        $-\frac{3 \beta_0}{\sqrt{2}}$ & 0  & 0 & 0 \\
        0 & 0 & $-\frac{\kappa_3}{2}$ & $-\frac{1}{4 \sqrt{2}}$ 
    \end{tabular}
    \right].
\end{aligned}
\end{equation}
In the following, we perform a detailed analysis of the Lagrangian
$\underline{L}_\Phi$.

\section{Scalar sector analysis}\label{sec:ScalarSector}

The Lagrangian $\underline{L}_\Phi$ for the scalar sector is rather
nontrivial. Our strategy here is as follows: we first integrate out the
auxiliary degrees of freedom, then diagonalize the Kinetic matrix of the
resulting system. The massive degrees of freedom are then identified
and integrated out.

\subsection{Integrate out nondynamical degrees of freedom}

We integrate out the nondynamical degrees of freedom $\Omega$ and $U$ by varying $\underline{L}_\Phi$ with respect to the same variables. We obtain the expressions:
\begin{equation} \label{Eq:IOAux}
    \begin{aligned}
        \Omega &= \frac{\sqrt{2} \beta_0 \left(3 \sqrt{X_0} \dot\psi-k^2 \phi\right)-\Theta}{2 \nu_1},\\
        U &= \frac{3\lambda}{4 \kappa_4} 
        \biggl[\sqrt{2} \eta k^2 \xi
        +2\sqrt{X_0} \left(4 \kappa_3 \dot\phi+\sqrt{2} \dot\Theta\right)+\frac{8 k^2 \kappa_1}{3\lambda} \psi
        \biggr].
    \end{aligned}
\end{equation}
Inserting these expressions back into $\underline{L}_\Phi$, we obtain a
reduced Lagrangian of the form:
\begin{equation} \label{Eq:PerturbedLagrangianScalarSectorFormIO}
    \begin{aligned}
        L_\Phi=&
            \dot{Y}^T \cdot {\mathbb{K}} \cdot \dot{Y}
            +
            {Y}^T \cdot {\mathbb{V}} \cdot {Y}
            +
            \dot{Y}^T \cdot {\mathbb{M}}^T \cdot {Y}.
    \end{aligned}
\end{equation}
where the the Kinetic matrix ${\mathbb{K}}$ and and the mass matrix
${\mathbb{V}}$ have the block diagonal form:
\begin{equation} \label{LagIOMatxs}
    \begin{aligned}
    {\mathbb{K}}&=
    \left[
    \begin{tabular}{cccc}
        ${K}_{11}$ & ${K}_{12}$ & 0 & 0 \\
        ${K}_{12}$ & 0 & 0 & 0 \\
        0 & 0 & ${K}_{33}$ & ${K}_{34}$ \\
        0 & 0 & ${K}_{34}$ & ${K}_{44}$  
    \end{tabular}
    \right]
    ,\quad
    {\mathbb{V}}=
    \left[
    \begin{tabular}{cccc}
        ${V}_{11}$ & ${V}_{12}$ & 0 & 0 \\
        ${V}_{12}$ & $V_{22}$ & 0 & 0 \\
        0 & 0 & ${V}_{33}$ & ${V}_{34}$ \\
        0 & 0 & ${V}_{34}$ & ${V}_{44}$  
    \end{tabular}
    \right],
\end{aligned}
\end{equation}
and the friction matrix ${\mathbb{M}}$ has the block form:
\begin{equation} \label{LagIOMatxsFriction}
    \begin{aligned}
    {\mathbb{M}}&=
    \left[
    \begin{tabular}{cccc}
        0 & 0 & ${M}_{13}$ & ${M}_{14}$ \\
        0 & 0 & ${M}_{23}$ & ${M}_{24}$ \\
        0 & 0 & 0 & 0 \\
        0 & 0 & 0 & 0  
    \end{tabular}
    \right]
    .
\end{aligned}
\end{equation}
The explicit expressions for the matrix components are given in Table
\ref{Tab:MatrixCoefficients}. It is worth pointing out here that both
${K}_{33}$ and ${V}_{33}$ are proportional to $k^4$ and that all
off-diagonal matrix components with the index label $3$ are proportional
to $k^2$; one may then absorb a factor of $k^2$ into a rescaling of
$Y_3=\phi$.

\subsection{Determinant of equation of motion matrix}
Before proceeding, it is perhaps worth considering the nature of the degrees of freedom in $L_\Phi$. Following \cite{Aoki:2019snr,Mukohyama:2018obj}, one may construct a matrix ${\mathbb{E}}$ of the form
\begin{equation} \label{Eq:EoMMatrix}
    \begin{aligned}
        {\mathbb{E}} = \omega^2 {\mathbb{K}} + 
            i \omega ({\mathbb{M}}-{\mathbb{M}}^T)
            +{\mathbb{V}} ,
    \end{aligned}
\end{equation}
such that if the components of $Y$ may be written as $Y_a = Y_{0,a} e^{i \omega_a t}$, the equation of motion for the system may be written
\begin{equation} \label{Eq:EoMForm}
    {\mathbb{E}} \cdot Y = 0.
\end{equation}
Since the dispersion relations form the eigenvalues of ${\mathbb{E}}$, the determinant $\det({\mathbb{E}})$ is proportional to the product of the dispersion relations for $Y$, as well as an additional overall factor $k^4$, which (as discussed earlier) can be absorbed into a rescaling of $Y_3=\phi$. If the dispersion relations are analytic at $k^2=0$, the determinant may be assumed to have the form (where $f_a(0)=0$, and $a\in\{1,2,3,4\}$):
\begin{equation} \label{Eq:EoMMatrixDet}
    \begin{aligned}
        \det({\mathbb{E}}) 
        \propto 
        \prod^4_a \left[\omega^2 - G_a k^2 - m_a^2 + f_a(\omega^2) k^2 + \mathcal{O}(k^4) \right] = 0,
    \end{aligned}
\end{equation}
which is a general expression valid for low momenta.\footnote{That the dispersion relations can be expanded in $k^2$ follows from the fact only even powers of $k$ appear in the components of ${\mathbb{E}}$.} Accounting for the fact that $\det({\mathbb{E}}) \propto k^4$, one can obtain the masses from Eq. \eqref{Eq:EoMMatrix} by taking the limit $k^2 \rightarrow 0$, which yields an expression of the form:
\begin{equation} \label{Eq:EoMMatrixDetMassFactoring}
    \begin{aligned}
        &\lim_{k^2\rightarrow 0}\frac{\det({\mathbb{E}})}{k^4}
        \propto 
        \omega^2
        \left(\omega^2-m_2^2\right)
        \left(3\eta \nu_1 \omega^4 + \mu_6 \omega^2 -4 X_0 \mu_7\right),
    \end{aligned}
\end{equation}
which implies the following for the masses:
\begin{equation} \label{Eq:Masses}
    \begin{aligned}
        m_1^2&=0 , \qquad m_2^2=-\lambda \mu_3, \\
        m_3^2&=-\frac{12 X_0 \eta+\mu_5 + \sqrt{(12 X_0 \eta+\mu_5)^2-48X_0\eta\nu_1\mu_7}}{6\eta \nu_1},\\
        m_4^2&=-\frac{12 X_0 \eta+\mu_5 - \sqrt{(12 X_0 \eta+\mu_5)^2-48X_0\eta\nu_1\mu_7}}{6\eta \nu_1}.
    \end{aligned}
\end{equation}
Now, since the masses are, in general, different, the dispersion relations will, in general, be different. Given the ansatz used in Eq. \eqref{Eq:EoMMatrixDet} for the dispersion relations, one can construct a recursion relation and expand in $k^2$ to obtain the dispersion relations of the form:
\begin{equation} \label{Eq:DispersionRelationRec}
    \begin{aligned}
        \omega^2 =& \, m_a^2 + {G}'_a  k^2 + \mathcal{O}(k^4),\\
        {G}'_a :=&  \, G_a - f_a(m_a^2)
    \end{aligned}
\end{equation}
If the masses are different, one may then obtain an expression for the effective gradient factor ${G}'_a$ for the massless degree of freedom by inserting the expression \eqref{Eq:DispersionRelationRec} for a given $m_a^2$ into the determinant equation $\det(\bar{\mathbb{E}})=0$, expanding to $\mathcal{O}(k^2)$, then solving for $G_a$. The result is:
\begin{equation} \label{Eq:GradientFactors}
    \begin{aligned}
        G'_1&= G_1= -\frac{2X_0 \mu_2 \nu_1}{\mu_3\mu_7}, \qquad
        G'_2= -\frac{3 Z+ X_0(2\beta_0+\gamma_0)}{3\mu_3},\\
        G'_3&= -\frac{12 \nu_1 \left(3 \eta  m_3^6-\lambda  \mu_2X_0^2\right)+m_3^2 (\mu_s+\mu_9 m_3^2)}{3 m_3^2 (m_3^2+\lambda\mu_3) (\mu_5+6 \eta  \nu_1 m_3^2+12 \eta X_0)},\\
        G'_4&= -\frac{12 \nu_1 \left(3 \eta  m_4^6-\lambda  \mu_2X_0^2\right)+m_4^2 (\mu_s+\mu_9 m_4^2)}{3 m_4^2 (m_4^2+\lambda\mu_3) (\mu_5+6 \eta  \nu_1 m_4^2+12 \eta X_0)},
    \end{aligned}
\end{equation}
where $G'_3$ and $G'_4$ have the same form, differing only in the masses.
Recall conditions \eqref{Eq:ConditionGhostFree0} and \eqref{Eq:ConditionLorentzianTensor}, which can be respectively written as $\mu_3>0$ and $\mu_2\mu_3>0$. We find that in order for the massless scalar degree of freedom to satisfy a Lorentzian dispersion relation (so that $G_1>0$), one must require $\mu_7<0$. However, we will consider both signs for $\mu_7$ in the following analysis.

To see whether there exists a suitable parameter space, we here study some specific cases. For this purpose, we consider a parameterization of the form:
\begin{equation} \label{Eq:Parameterization}
    \begin{aligned}
        X_0=q Z/\gamma_0,\quad\beta_0=\frac{2(1+q)\gamma_0}{ql}-3\alpha_0.
    \end{aligned}
\end{equation}
The condition $\mu_2>0$ requires that $q>1$, and $\mu_7<0$ requires $l>1$ (assuming the other constants of the theory are positive). It is not too difficult to verify by direct computation that for $\lambda=\eta=\alpha_0=1$, $q=\gamma_0=3$, and $l=2$, one has 
\begin{equation} \label{Eq:Negativemu7}
    \begin{aligned}
        G'_1&=1/5,\quad
        G'_2=-8/15,\quad
        G'_3=-1.60,\quad
        G'_3=-2.07,\\
        m_2^2&=-5Z, \quad
        m_3^2=-2.92 Z,\quad
        m_4^2=-0.913 Z.
    \end{aligned}
\end{equation}
Now in the case of positive $l<1$, one has $\mu_7>0$, so that $G'_1<0$. In this case, one can seek to ensure that the dispersion relations for the remaining scalar degrees of freedom are Euclidean and the masses are tachyonic. For $\eta=-1$, $\lambda=\gamma_0=\alpha_0=1$, $q=4$ and $l=1/3$, we obtain:
\begin{equation} \label{Eq:Positivemu7}
    \begin{aligned}
        G'_1&=-33/230,\>\,
        G'_2=-43/69,\>\,
        G'_3=-1/2,\>\,
        G'_4=-2.73,\\
        m_2^2&=-23Z, \quad\>\,\,
        m_3^2=-4 Z,\quad\>\,\,
        m_4^2=-4.85 Z.
    \end{aligned}
\end{equation}
One can verify that in each case, there exists a neighborhood of parameter space around these points such that $G'_2$, $G'_3$, $G'_4$ are all negative, and the nonvanishing masses are tachyonic. Recalling our earlier discussion of the vector modes, we emphasize again that Euclidean tachyonic modes can be easily controlled
by suitable boundary conditions.

\subsection{Integrating out degrees of freedom}
To formulate a strategy for integrating out degrees of freedom, it is perhaps appropriate to consider the behavior of the matrices $\mathbb{K}$, $\mathbb{V}$, and $\mathbb{M}$ for small $k^2$, only keeping terms to leading order in $k^2$:
\begin{equation} \label{LagIOMatxKsk}
    {\mathbb{K}}\approx
    \left[
    \begin{tabular}{cccc}
        $\mathcal{K}$ & $-\tfrac{3\eta}{4\sqrt{2}}$ & 0 & 0 \\
        $-\tfrac{3\eta}{4\sqrt{2}}$ & 0 & 0 & 0 \\
        0 & 0 & $\tfrac{k^4}{3X_0\lambda}$ & $\tfrac{k^4}{12\sqrt{2}X_0^2\lambda}$ \\
        0 & 0 & $\tfrac{k^4}{12\sqrt{2}X_0^2\lambda}$ & $-\tfrac{1}{16X_0}$  
    \end{tabular}
    \right]
\end{equation}
\begin{equation} \label{LagIOMatxVsk}
    {\mathbb{V}}\approx
    \left[
    \begin{tabular}{cccc}
        $k^2 \mu_2$ & $-\tfrac{k^2\eta}{2\sqrt{2}}$ & 0 & 0 \\
        $-\tfrac{k^2\eta}{2\sqrt{2}}$ & $\tfrac{\eta}{16}$ & 0 & 0 \\
        0 & 0 & $\tfrac{k^4\beta_0\nu_2}{2\nu_1}$ & $\tfrac{k^2\alpha_0}{2\sqrt{2}\nu_1}$ \\
        0 & 0 & $\tfrac{k^2\alpha_0}{2\sqrt{2}\nu_1}$ & $-\tfrac{1}{4\nu_1}$  
    \end{tabular}
    \right],
\end{equation}
\begin{equation} \label{LagIOMatxMsk}
    {\mathbb{M}}\approx
    \left[
    \begin{tabular}{cccc}
        0 & 0 & $\tfrac{k^2 \mathcal{K}}{3\sqrt{X_0}}$ & $\tfrac{3\alpha_0\sqrt{X_0}}{2\sqrt{2}\nu_1}$ \\
        0 & 0 & $-\tfrac{k^2\eta}{4\sqrt{2}\sqrt{X_0}}$ & $-\tfrac{k^2 \eta}{32 X_0^{3/2}}$ \\
        0 & 0 & 0 & 0 \\
        0 & 0 & 0 & 0
    \end{tabular}
    \right].
\end{equation}
The dependence on $k^2$ in the respective Kinetic and mass matrices $\mathbb{K}$ and $\mathbb{V}$ indicates that $Y_3=\phi$ is scaled by a factor of $k^2$, and that $Y_1=\psi$ is connected to the massless degree of freedom for small $k^2$. From the form of $\mathbb{V}$, we expect the massive degrees of freedom to be connected to $Y_2=\xi$ and $Y_4=\Theta$.

We, therefore, begin by integrating out the massive degrees of freedom $\xi$ and $\Theta$; we do this by varying $L_\Phi$ with respect to $\xi$ and $\Theta$, setting second derivatives to zero in the resulting equations of motion, and solving $\xi$ and $\Theta$. We then resubstitute back into the action (setting $\dot\xi=0$ and $\dot\Theta=0$) to obtain the Lagrangian. The resulting expression is:
\begin{equation} \label{LpsiExpanded03}
L'_\Phi = 
K'_{\psi\psi} \dot{\psi}^2+
K'_{\underline{\phi}\underline{\phi}} \dot{\underline{\phi}}^2+
V'_{\underline{\phi}\underline{\phi}} \underline{\phi}^2 + V'_{\psi\psi} \psi^2+M'_{\underline{\phi}\psi}\underline{\phi}\dot{\psi},
\end{equation}
where $\underline{\phi}:=\phi k^2$, and the coefficients are given by some rather complicated expressions:
\begin{equation} \label{LpsiExpanded03coeffs}
    \begin{aligned}
    K'_{\underline{\phi}\underline{\phi}}
    &= \frac{\kappa_3(2-3\eta\lambda)}{2\kappa_4-3k^4\eta\lambda},
    \qquad\qquad
    V'_{\underline{\phi}\underline{\phi}}
    = \frac{\nu_1}{2},
    \\
    K'_{\psi\psi} 
    &= \frac{3X_0\lambda(X_0\kappa_3-k^2\mu_1)(2\kappa_4+3\lambda(k^2\mu_1-X_0\kappa_3))\nu_1}{2\kappa_4^2} \\
    &\quad  + \bar\mu_{{\rm (4,1,-3,-3)}},
    \\
    V'_{\psi\psi}
    &= \frac{k^2}{2\kappa_4-3k^4\eta\lambda}\biggl[12\lambda X_0^2\mu_2 -9\eta k^6 \\
    &\qquad\qquad\qquad\quad\>\>
    -(3\eta\lambda \bar\mu_{{\rm (0,4,5,3)}}-2\bar\mu_{\rm (0,-1,1,3)})k^4\\
    &\qquad\qquad
    +(4X_0^2(1+3\lambda(\beta_0\gamma_0-2\eta))-6\lambda \mu_1\mu_3)k^2 \biggr],\\
    M'_{\underline{\phi}\psi}
    &= \frac{\sqrt{X_0}}{2\kappa_4-3k^4\eta\lambda}\biggl[
    6\lambda\kappa_3\mu_7 - 8 (1+3\eta\lambda) X_0 k^2
    \\
    &\qquad\qquad\qquad\quad- (\alpha_0(9\eta\lambda-6)+\beta_0(15\eta\lambda+2)) k^4\biggr].
    \end{aligned}
\end{equation}
The appearance of quadratic terms in $\underline{\phi}$ when $k^2\rightarrow 0$ indicates that $\underline{\phi}$ is connected to a massive degree of freedom, so we integrate out $\underline{\phi}$ as well. After integrating out $\underline{\phi}$, we obtain the following Lagrangian, keeping terms up to $\mathcal{O}(k^2)$:
\begin{equation} \label{LpsiExpanded02}
    L_\psi = \frac{\mu_3\mu_7}{X_0\nu_1} \dot\psi^2 + k^2 \left(\mu_2\psi^2+\frac{2(1+3\eta\lambda) \mu_7}{3X_0 \nu_1 \lambda}\dot\psi^2\right).
\end{equation}
One can see that $\psi$ is massless, and it is not too difficult to confirm that $\psi$ indeed satisfies the massless dispersion relation in Eq. \eqref{Eq:GradientFactors}. Now recall that conditions \eqref{Eq:ConditionGhostFree0} and \eqref{Eq:ConditionLorentzianTensor} may be respectively written as $\mu_3>0$ and $\mu_3\mu_2>0$, and correspond to the requirement that the tensor modes  $h$ have a positive kinetic term and satisfy a Lorentzian dispersion relation. If $\psi$ also satisfies a Lorentzian dispersion relation, one requires that $\mu_7<0$, but it follows that the kinetic term for $\psi$ is negative valued. More generally, if $h$ and $\psi$ both satisfy Lorentzian dispersion relations, either $h$ or $\psi$ is necessarily ghostlike.\footnote{It should be mentioned that there are some scenarios that can, in principle, accommodate ghosts---see for instance, \cite{Deffayet:2025lnj,*Deffayet:2023wdg,*Deffayet:2021nnt} and \cite{Hell:2023rbf,*Maldacena:2011mk}.}

On the other hand, we can instead consider $\mu_7>0$, in which case $\psi$ satisfies a massless Euclidean dispersion relation. If we assume that $k^2$ is small compared to the factor ${\mu_3\mu_7}/{X_0\nu_1}$ (of order $M_\mathrm{Pl}^2$), we obtain:
\begin{equation} \label{LpsiExpanded02appx}
    L_\psi \approx \frac{\mu_3\mu_7}{X_0\nu_1} \dot\psi^2 + k^2 \mu_2\psi.
\end{equation}
This indicates that under an appropriate rescaling, $\psi$ is a harmonic function for $\mu_7>0$, which is, in principle, bounded by its value on the boundary. Since the constants in $L_\psi$ have dimension $M_\mathrm{Pl}^2$, any contribution arising from higher order effects will, in principle, be suppressed\footnote{It is perhaps worth noting that in Eqs. (\ref{LagIOMatxKsk}--\ref{LagIOMatxMsk}), the only terms of mass dimension $M_\mathrm{Pl}^2$ or higher are those containing $\psi$.} by an additional factor of order $1/M_\mathrm{Pl}^2$, so that despite the absence of a large tachyonic mass, one can still suppress $\psi$ by setting its value on the boundary to zero.

\section{Summary and Discussion}\label{sec:Disc}
In this article, we considered the Euclidean signature scalar-tensor theory of \cite{Mukohyama:2013gra}, which is renormalizable \cite{Muneyuki:2013aba} and can evade Ostrogradskian instabilities \cite{Woodard:2015zca,*Woodard:2023tgb,*Ostrogradsky1850} by virtue of the absence of genuine dynamics. In this theory, the concept of dynamics emerges locally for physics at low momenta. At extremely high momenta, or short distances, there is no concept of dynamics as the theory is genuinely locally Euclidean. Globally, or at very long distances, again there is no dynamics as everything is in principle determined by suitable boundary conditions. In particular, we conducted an analysis of perturbations over a flat background with a non-vanishing derivative of the clock field and their associated dispersion relations with the aim of identifying massless degrees of freedom with Lorentzian dispersion relations. 

We find in our analysis a massless tensor degree of freedom that can, with an appropriate choice of parameters, satisfy a Lorentzian dispersion relation. At the same time, we show that all but one of the remaining degrees of freedom can be made to satisfy Euclidean dispersion relations with a large tachyonic mass; these modes can be set to zero with an appropriate choice of boundary values. The last degree of freedom is a massless scalar that can either be a massless ghost satisfying a Lorentzian dispersion relation or a harmonic function. We have argued that the latter is preferable, as harmonic functions are bounded by their values on the boundary, and higher order effects are suppressed by $M_\mathrm{Pl}^2$. In this manner, one can argue that the degrees of freedom that do not satisfy Lorentzian dispersion relations can be suppressed.

A next step in this direction will be to extend the present analysis to different backgrounds, such as those corresponding to Lorentz signature black holes; it is important to confirm (as one might expect) that conclusions of the present analysis remain valid so long as the momentum scale is larger than the (intrinsic and extrinsic) curvature scale, and much smaller than $M_\mathrm{Pl}$. It may be of interest to study the fate of modes with momenta comparable to or lower than the curvature scale in such backgrounds. One can also extend the analysis to include a simple matter model, and then all ingredients necessary to accommodate the standard model of particle physics \cite{Kehayias:2014uta}. If one can recover a Lorentzian dispersion relation for matter degrees of freedom, one might expect additional constraints to arise in such an analysis from the requirement that the massless tensor and matter degrees of freedom propagate with the same effective sound speed, thus recovering Lorentz invariance. As discussed earlier, it is perhaps appropriate in such future studies to consider a careful study of the renormalization group flow of the theory, and to also adopt the idea of the emergent Lorentz invariance \cite{Chadha:1982qq}.

%
%


\begin{acknowledgments}
JCF thanks Will Barker, Ignacy Sawicki and Atabak Jalali for helpful discussions and reference suggestions, and acknowledges the Leung Center for Cosmology and Particle Astrophysics (LeCosPA), National Taiwan University (NTU), the R.O.C. (Taiwan) National Science and Technology Council (NSTC) Grant No. 112-2811-M-002-132, which supported part of this work.  JCF is also supported by the European Union and Czech Ministry of Education, Youth and Sports through the FORTE project No. CZ.02.01.01/00/22\_008/0004632.
The work of SM was supported in part by Japan Society for the Promotion of Science (JSPS) Grants-in-Aid for Scientific Research No.~24K07017 and the World Premier International Research Center Initiative (WPI), MEXT, Japan. The work of SC has been carried out in the framework of activities of the INFN Research Project QGSKY.

\end{acknowledgments}


%
%


\bibliography{ref}

\end{document}